\documentstyle[twoside,fleqn,espcrc2]{article}

\title{ 
\hfill
\begin{minipage}{0pt}\scriptsize\vspace*{-1.5cm} \begin{tabbing}
\hspace*{\fill} HD--THEP--99--38
\end{tabbing} 
\end{minipage}\\[-8pt]
Gauge--invariant nonlocal quark condensates in QCD:\\
a new interpretation of the lattice results}

\author{E. Meggiolaro\address{Institut 
         f\"ur Theoretische Physik, Universit\"at Heidelberg, 
         Philosophenweg 16, D--69120 Heidelberg, Germany}}
\begin{document}

\begin{abstract}

We study the asymptotic short--distance behaviour as well as
the asymptotic large--distance behaviour of the gauge--invariant
quark--antiquark nonlocal condensates in QCD.
A comparison of some analytical results with the
available lattice data is performed.

\end{abstract}

\maketitle

\section{INTRODUCTION}

In a recent paper~\cite{paper99} we have presented a lattice determination
of the quark--antiquark nonlocal condensates, defined as:
\begin{equation}
C_i (x) = -\displaystyle\sum_{f=1}^4
\langle {\rm Tr} [\bar{q}^f_a (0) (\Gamma^i)_{ab} S(0,x) q^f_b (x)] \rangle ~,
\end{equation}
where $S(0,x)$ is the Schwinger line
needed to make $C_i (x)$ gauge--invariant and  $\Gamma^i$ are the
sixteen independent  matrices of the Clifford's algebra
acting on the Dirac indices $a,b$.
The trace in (1) is taken with respect to the colour indices.

Making use of T,P invariance one 
can prove that all the correlators (1) vanish, except 
those with $\Gamma^i = {\bf 1}$ (``scalar'' nonlocal condensate)
and with  $\Gamma^i = \gamma_E^\mu$ and $\mu$ in the direction of $x$ 
(``longitudinal--vector'' nonlocal condensate):
\begin{eqnarray}
C_0 (|x|) = -\displaystyle\sum_{f=1}^4 \langle {\rm Tr} [ \bar{q}^f_a (0)
S(0,x) q^f_a (x) ] \rangle ~; \nonumber \\
C_v (|x|) =
-{x_\mu \over |x|}  \displaystyle\sum_{f=1}^4 \langle {\rm Tr} [ \bar{q}^f_a 
(0) (\gamma^\mu_E)_{ab} S q^f_b (x) ] \rangle ~.
\end{eqnarray}
These two quantities play a relevant role in many applications of QCD sum 
rules, especially for studying the meson form factors and the meson wave 
functions~\cite{Mik-Rad86-92,Rad91,Bak-Rad91,Bak-Mik98}.
The nonlocal quark condensates have been also determined within the
single instanton approximation of the instanton liquid model 
~\cite{Dor-Esa-Mik97}.

The lattice computations of Ref.~\cite{paper99} have been performed 
both in the {\it quenched} approximation and in full QCD using 
four degenerate flavours of {\it staggered}
fermions [whence the sum over the flavour index $f$ in (1)]
and the $SU(3)$ Wilson action for the pure--gauge sector.

In full QCD the nonlocal condensates have been measured on a $16^3 \times 24$ 
lattice at $\beta = 5.35$ and two different values of the quark mass: 
$a \cdot m_q = 0.01$ and $a \cdot m_q = 0.02$ ($a$ being the lattice spacing). 
For the {\it quenched} case the measurements have been performed on a $16^4$
lattice at $\beta = 6.00$, using valence quark masses $a \cdot m_q = 0.01$,
$0.05$, $0.10$, and at $\beta = 5.91$ with a  quark mass $a \cdot m_q = 0.02$. 
Further details, as well as a remark about the reliability of the results 
obtained for the longitudinal--vector nonlocal condensate, can be found 
in~\cite{paper99}.

In what follows we shall concentrate on the scalar nonlocal condensate
$C_0 (x)$. In Ref.~\cite{paper99} a best fit to the data with the following 
function has been tried:
\begin{equation}
C_0 (x) = A_0 \exp (-\mu_0 x) + {B_0 \over x^2} ~.
\end{equation}
The form of the perturbative--like term $B_0/x^2$ is that obtained in the 
leading order in perturbation theory, in the chiral limit $m_q \to 0$. 

The quantity of greatest physical interest which can be extracted from our 
lattice determinations is the correlation length $\lambda_0 \equiv 1 / \mu_0$
of the scalar quark--antiquark nonlocal condensate. 
At the lightest quark mass $a \cdot m_q = 0.01$ we have obtained the value 
$\lambda_0 \simeq 0.63$ fm ~\cite{paper99}. 

Here we study the asymptotic short--distance behaviour as well as
the asymptotic large--distance behaviour of the gauge--invariant
quark--antiquark nonlocal condensates in QCD.
The large--distance behaviour is
derived by making use of a relation of these correlators to the
two--point functions for the scalar and pseudoscalar $\overline{Q} q$
meson operators in the limit of the heavy--quark mass $M_Q \to \infty$.
A comparison of some analytical results with the
available lattice data will be performed.

\section{SHORT--DISTANCE BEHAVIOUR}

\noindent
The short--distance behaviour of the correlators is described by an
``operator product expansion'' (O.P.E.), having the form:
\begin{equation}
C_0(x) = {\tilde{B}_0 \over x^2} - \langle : \bar{q} q : \rangle + \ldots
\end{equation}
The vacuum expectation values of the local operators, such as
$\langle : \bar{q} q : \rangle$, appear as expansion coefficients of the
nonlocal condensate $C_0(x)$ in a Taylor series in the variable $x^2$.
In the literature~\cite{Mik-Rad86-92,Rad91,Bak-Rad91}
the nonperturbative part of the scalar nonlocal condensate is often 
parametrized with a Gaussian function:
\begin{equation}
C^{(n.p.)}_0(x) = \tilde{A}_0 \exp \left( -{1 \over 8} \mu_q^2 x^2 \right) ~.
\end{equation}
The parameter $\mu_q^2$ characterizes the nonlocality of the quark
condensate and it is given by 
~\cite{Mik-Rad86-92,Rad91,Bak-Rad91}:
\begin{equation}
\mu_q^2 = {\langle : \bar{q} D^2 q : \rangle \over
\langle : \bar{q} q : \rangle} ~,
\end{equation}
where $D_\mu = \partial_\mu +i g A^a_\mu T^a$ is the covariant derivative.
By the equations of motion in the chiral limit, the parameter $\mu_q^2$
is also related to the mixed quark--gluon condensate:
\begin{equation}
\mu_q^2 = {\langle : \bar{q} (i g \sigma_{\mu\nu} G^a_{\mu\nu} T^a) q : 
\rangle \over 2 \langle : \bar{q} q : \rangle} ~.
\end{equation}
From the QCD sum--rules phenomenology one finds the following estimate
for $\mu_q^2$~\cite{Pivoravov91}:
\begin{equation}
\mu_q^2 = 0.50(5) ~{\rm GeV}^2 ~.
\end{equation}
Therefore we have also tried a best fit to the data of the scalar
nonlocal condensate with the function
\begin{equation}
C_0(x) = \tilde{A}_0 \exp \left( -{1 \over 8} \mu_q^2 x^2 \right) 
+ {\tilde{B}_0 \over x^2} ~.
\end{equation}
It comes out that the data are well fitted by this function [even if
the $\chi^2/N_{d.o.f.}$ is slightly larger than in the exponential
case, Eq. (3)]. The following value for $\mu_q^2$ is obtained from the 
best fit to the full--QCD data at $\beta = 5.35$ and $a \cdot m_q = 0.01$:
\begin{equation}
\mu_q^2 = 0.46(5) ~{\rm GeV}^2 ~.
\end{equation}
This value is in good agreement with the QCD sum--rule estimate (8).

\section{LARGE--DISTANCE BEHAVIOUR}

\noindent
In this section we study the asymptotic large--distance behaviour of
the quark correlators.
The starting point is the fermion propagator in an external gluon
field $A_\mu = A^a_\mu T^a$, in the infinite mass limit $M_Q \to \infty$
(static fermion limit) ~\cite{Brown-Weisberger79,Eichten-Feinberg81}:
\begin{eqnarray}
\lefteqn{
\langle Q_{a,\alpha}(z) \overline{Q}_{b,\beta}(z') \rangle_{(A)} = }
\nonumber \\
& & = \delta^{(3)}(\vec{z} - \vec{z}')
\left[ {\rm P} \exp \left( i g \displaystyle\int_{z_0}^{z'_0} {\rm d}t
A_0 (\vec{z},t) \right) \right]_{\alpha\beta} \nonumber \\
& & \times \left[ \theta (z_0 - z'_0) ({\rm P}_+)_{ab}
{\rm e}^{-i M_Q (z_0 - z'_0)} \right. \nonumber \\
& & + \left. \theta (z'_0 - z_0) ({\rm P}_-)_{ab}
{\rm e}^{i M_Q (z_0 - z'_0)} \right] ~,
\end{eqnarray}
where $\alpha,\beta = 1, \ldots, N_c$ are colour indices, $a,b$ are Dirac 
indices and
\begin{equation}
{\rm P}_+ \equiv {{\bf 1} + \gamma^0 \over 2} ~~~ ; ~~~
{\rm P}_- \equiv {{\bf 1} - \gamma^0 \over 2} ~.
\end{equation}
Let us consider now the following mesonic correlators:
\begin{eqnarray}
M^{(Q)}_S (x^0) \equiv \displaystyle\sum_{f=1}^4 \int {\rm d}^3 \vec{z}'
\langle O^{f\dagger}_S (x^0,\vec{z}') O^f_S (0) \rangle ~; \nonumber \\
M^{(Q)}_{PS} (x^0) \equiv \displaystyle\sum_{f=1}^4 \int {\rm d}^3 \vec{z}'
\langle O^{f\dagger}_{PS} (x^0,\vec{z}') O^f_{PS} (0) \rangle ~,
\end{eqnarray}
where $x^0 > 0$ and the mesonic operators $O^f_S$ and $O^f_{PS}$ are 
so defined:
\begin{equation}
O^f_S \equiv \bar{q}^f \gamma^0 Q ~~~ ; ~~~ 
O^f_{PS} \equiv \bar{q}^f \gamma^0 \gamma^5 Q ~.
\end{equation}
The spin and colour indices are contracted. In other words, $O^f_S$ is the
time component of the vector current $J^\mu_f = \bar{q}^f \gamma^\mu Q$,
while $O^f_{PS}$ is the time component of the axial--vector current
$J^\mu_{5,f} = \bar{q}^f \gamma^\mu \gamma^5 Q$: we call $O^f_S$
``scalar'' meson operator and $O^f_{PS}$ ``pseudoscalar'' meson operator.
Using Eq. (11) and performing the analytic continuation from
Minkowskian to Euclidean space ($x^0 \to -i x_4$, with $x^0,~x_4 > 0$),
one easily finds the following relations:
\begin{eqnarray}
M^{(Q)}_{E,PS} (x_4) - M^{(Q)}_{E,S} (x_4) = {\rm e}^{-M_Q x_4}
C_0 (x_4) ~; \nonumber \\
M^{(Q)}_{E,PS} (x_4) + M^{(Q)}_{E,S} (x_4) = {\rm e}^{-M_Q x_4}
C_v (x_4) ~.
\end{eqnarray}
$C_0 (x_4)$ and $C_v (x_4)$ are precisely the quark correlators measured
on the lattice, defined by Eqs. (2).
As a candidate for the heavy quark $Q$, we can take the $b$ ({\it bottom})
quark, for which we know the pseudoscalar mesons of the 
type $I(J^P) = {1 \over 2} (0^-)$, with mass $M_B \simeq 5.3$ GeV.
We now consider the Eqs. (15) in the asymptotic limit $x_4 \to \infty$.
If we make the assumption (supported by the experimental evidence) that
the lightest $B$--mesons are the pseudoscalar ones, we find that:
\begin{equation}
C_{0,v} (x_4) \mathop\sim_{x_4 \to \infty} A_\infty 
{\rm e}^{-\mu_\infty x_4} ~,
\end{equation}
where
\begin{equation}
A_\infty = 2 F_B^2 M_B ~~~ ; ~~~ \mu_\infty = M_B - M_b ~.
\end{equation}
$M_b$ is the mass of the $b$ quark.
$F_B$ is the $B$--meson decay constant, defined in the usual way:
\begin{equation}
\langle B(\vec{p}) | J^\mu_{5,f} (x) | 0 \rangle = -i F_B p^\mu
{\rm e}^{i p x} ~.
\end{equation}
We have indicated with $| B(\vec{p}) \rangle$ the state of a $B$--meson
with spatial momentum $\vec{p}$. This state is normalized in the
following way:
\begin{equation}
\langle B(\vec{p}) | B(\vec{q}) \rangle = 2 p^0_B (2\pi)^3
\delta^{(3)} (\vec{p} - \vec{q}) ~,
\end{equation}
where $p^0_B = \sqrt{\vec{p}^2 + M_B^2}$ is the energy.

Using the above--mentioned value of $M_B$ and the rough estimate
$M_b \approx 5$ GeV for the mass of the $b$ quark, we derive:
\begin{equation}
\mu_\infty = M_B - M_b \approx 300 ~{\rm MeV} ~.
\end{equation}
That is, after conversion to length units,
$\lambda_\infty \equiv 1 / \mu_\infty \approx 0.66$ fm.
This value is in good agreement with the value $\lambda_0 \simeq 0.63$ fm 
that we have obtained at the lightest quark mass $a \cdot m_q = 0.01$
from a best fit with the function (3).
An approach which uses the QCD sum--rules techniques gives the value
$\mu_\infty \approx 290 \div 360$ MeV \cite{Dosch99}, which is in good 
agreement with the estimate (20) and with the above--mentioned lattice result.

\section{DISCUSSION AND CONCLUSIONS}

\noindent
We have reconsidered the available lattice data for the gauge--invariant
quark--antiquark nonlocal condensates~\cite{paper99}.
A gaussian--type parametrization, inspired by the O.P.E. at short 
distances, fits well the data and gives results in agreement with 
the QCD sum--rules phenomenology.
The original exponential--type parametrization fits also very well the data,
in agreement with the expected large--distance behaviour of the
correlators.
The natural question which arises is: are the available lattice data for the 
correlators in the short--distance or in the large--distance regime?
It seems that they are just in an intermediate range.
Further study is required in order to investigate more accurately
both the short--distance and the large--distance asymptotic behaviour.

\section*{Acknowledgements}
This work was supported by the EC TMR program ERBFMRX--CT97--0122.
I would like to thank A. Di Giacomo and H.G. Dosch for many useful
discussions.

\end{document}